\documentclass[aps,graphicx,twocolumn,epsfig]{revtex4}

\usepackage{epsfig}
\usepackage{amsmath}

\newcommand{\ket}[1]{|#1\rangle}

\begin{document}
 
\author{C. Trefzger$^{1}$} 
\author{C. Menotti$^{1,2}$}
\author{M. Lewenstein$^{3}$} 
\affiliation{$^1$ ICFO - Institut de Ciencies Fotoniques, 
Mediterranean Technology Park, 08860 Castelldefels (Barcelona), Spain \\ 
$^2$ CRS BEC-INFM and Dipartimento di Fisica, Universit\`a di Trento, 
I-38050 Povo, Italy\\ 
$^3$ ICREA - Instituci\'o Catalana de Recerca i Estudis Avan\c cats and  ICFO - Institut de Ciencies Fotoniques, 
Mediterranean Technology Park, 08860 Castelldefels (Barcelona), Spain}

\title{Ultracold dipolar gas in an optical lattice: the fate of metastable states}

\begin{abstract}
We study the physics of ultracold dipolar bosons in optical lattices. We show that dipole-dipole interactions lead to the appearance of many insulating metastable states. We study the stability and lifetime of these states using a generalization of the instanton theory. We investigate also possibilities to prepare, control and manipulate these states using time dependent superlattice modifications and modulations. We show that the transfer from one metastable configuration to another necessarily occurs via superfluid states, but can be controlled fully on the quantum level. We show how the metastable states can be
created in the presence of the harmonic trap. Our findings open the way toward applications of  the metastable states as quantum memories.  
\end{abstract}

\maketitle

\section{Introduction}
Ultracold dipolar gases  have recently attracted a lot of attention, both from the experimental and theoretical side \cite{review1,review2,review3,review4}.
Experiments show that it is possible to cool Chromium below the degeneracy temperature \cite{pfau}, and by using a Feshbach resonance 
one can reduce the $s$-wave contact interaction \cite{Lahaye} such that the physics of the system is dominated by the anisotropic dipole-dipole interaction between atoms.
These achievements together with the rather remarkable progress in cooling and trapping of dipolar molecules \cite{EPJD} clearly indicates that  these systems indeed are within experimental reach. Particularly interesting in this context are ultracold dipolar gases in optical lattices \cite{review2,review5}, which offer novel possibilities of studying strongly correlated states. 

Theoretical studies \cite{Fisher,Jaksch}, and experiments \cite{Bloch} on bosons in optical lattices have pointed out the existence of two main kinds of phases: (i) a superfluid phase (SF), characterized 
by a uniform non-zero order parameter, in which particles are delocalized over the whole lattice, and (ii) a Mott-insulator phase (MI) of localized atoms, in which the
order parameter is zero. As soon as one introduces a long range interaction between atoms, new phases appear, both in the SF \cite{batrouni,sengupta} and MI \cite{sinha,Goral}
region of the phase diagram: the supersolid phase (SS), which features a non-zero order parameter following a modulated pattern, and the charge density wave (CDW) in which the 
localized atoms follow modulated patterns with a resulting zero order parameter. These effects are particularly strong in dipolar gases in optical lattices \cite{Goral}. Moreover, it has been shown \cite{Menotti} that the lower tunneling region of the 
phase diagram is characterized by the existence of many almost degenerate metastable states of Mott-like distribution of atoms in the lattice. These distributions 
consist of localized atoms following a specific pattern with a filling factor (average number of atoms per site) which is in general not integer. 

The large number of metastable states suggests the analogy between dipolar gases and classical complex systems such as neural network models \cite{Amit}, or spin glass models \cite{parisi}. As it is well known, classical complex systems, and in particular neural networks, may serve very well as very efficient classical distributed memory models. They are robust with respect to the damage of part of the network, and they "recognize patterns with distortion", acting as associative memory. It is natural to ask if these properties could not turn out to be useful also for quantum memories. Quantum memories serve to store in a robust way quantum states, i.e. not only some "classical" patterns, but also quantum fluctuations (see for instance
\cite{polzik}). It would be interesting to combine the best of the two worlds: robustness,  associativity
and large storage capacity of "classical" distributed memories, with quantum stability for the storage of fluctuations. This is the motivation and far reaching goal of this paper. The paper itself concentrates on the first steps toward this goal: stability, control, preparation and manipulation of the metastable states in ultracold dipolar gases in optical lattices.

We focus on the \textit{insulating} metastable states of the system. We show how the appearance of these states crucially depends on the dipole-dipole interaction. Our calculations on the stability of these states 
show that their lifetime strongly increases when
hopping is suppressed and scales exponentially with
the number of sites involved in the tunneling process
to other metastable states. We show that once the system is prepared in a certain metastable configuration, it is necessary 
to pass through the SF region of the phase diagram in order to dynamically pass from the given metastable configuration to another one, 
and that this is a quantum controlled process.

The paper is organized as follows. In Sect.~\ref{sec:MODEL} we introduce the model. In Sect.~\ref{sec:GS} we first derive the mean field (MF) Hamiltonian, then 
calculate the ground and metastable states of the system, for the case in which the dipole-dipole interaction is a small perturbation with respect to the contact interaction 
($U/U_{NN}$ = 20, see text for details). This section contains several subsections: we study the behavior of the system with respect to the cut-off range of the dipole-dipole interaction and the size of the elementary cell that reproduces the infinite lattice. Non-uniform lattices are also discussed. Low energy excitations are discussed in Sect.~\ref{sec:EXCITATIONS}, while the stability of the metastable states is 
discussed in Sect.~\ref{sec:STABILITY}, through an instanton approach. In Sect.~\ref{sec:DYNAMICS} we study how to manipulate, in a deterministic way, the metastable
configurations by changing in time the lattice parameters. In Sect.~\ref{sec:FINITE}, we treat the effects of a confining trap on the system.
We discuss our results in Sect.~\ref{sec:conclusion}.

\section{The model}
\label{sec:MODEL}
We study a single component gas of bosons (i.e. spin or pseudo-spin,
polarized) \cite{Goral,sengupta2} in an optical lattice. We assume the temperature of the system to be low enough
such that we can restrict to the first Bloch band, and the system is well described by the 
extended Bose-Hubbard Hamiltonian:
\begin{eqnarray}
H &=& -\frac{J}{2} \sum_{\langle i j\rangle}\left(
 a^{\dag}_i a_j + a_i a^{\dag}_j \right) -  \sum_i \mu n_i \nonumber\\ 
&+&\sum_i \frac{U}{2}  \; n_i(n_i-1) 
+  \sum_{\vec \ell} 
\sum_{\langle \langle i
 j \rangle \rangle_{\vec \ell}} \frac{U_{\vec \ell}}{2} \; n_i n_j ,
\label{GBH}
\end{eqnarray}
where $J$ is the tunneling coefficient, $U$ the on-site interaction,
$U_{\vec \ell}$ the strengths of the dipole-dipole interaction at
different relative distances, and $\mu$ the chemical potential which
fixes the average atomic density. In our notation $\langle i j \rangle$
represents nearest neighbors, and $\langle \langle i j \rangle
\rangle_{\vec \ell}$ represents neighbors at distance $\vec \ell$.

We describe our system with Hamiltonian (\ref{GBH}), and a Gutzwiller ansatz 
for the wave function \cite{Jaksch}
\begin{eqnarray}
|\Phi(t)\rangle= \prod_i \sum_n f_n^{(i)}(t) |i,n\rangle,
\label{GW}
\end{eqnarray}
where $|i,n\rangle$ denotes the Fock state of $n$ atoms at site $i$.
In particular the time-dependence of the Gutzwiller coefficients
$f_n^{(i)}$ allows to study the evolution of the state in real and
imaginary ($\tau = -it$) time \cite{Jaksch01,Goral}

\begin{eqnarray}
\label{TDGA}
i \frac{d\,f_n^{(i)}}{dt} &=& -J \left[ \bar{\varphi}_i \sqrt{n_i}
  f_{n-1}^{(i)} + \bar{\varphi}_i^* \sqrt{n_i+1} f_{n+1}^{(i)} \right] +   \\ 
&+& \left[\frac{U}{2} n_i(n_i-1) + \sum_{\vec \ell} U_{\vec \ell}\;
  \bar{n}_{i,{\vec \ell}} \; n_i -\mu n_i \right] f_n^{(i)}, \nonumber
\end{eqnarray}
with $\varphi_i= \langle \Phi | a_i | \Phi \rangle$, $\bar{\varphi}_i=
\sum_{\langle j \rangle_i} \varphi_j$, $n_i=\langle \Phi | a^\dag_i
a_i | \Phi \rangle$, and $\bar{n}_{i,{\vec \ell}} = \sum_{\langle
\langle j \rangle\rangle_{i,{\vec \ell}}} n_j$.

We consider an infinite two dimensional (2D) square lattice. We assume the dipoles to be polarized such that the atoms in the lattice experience a repulsive dipole-dipole interaction 
in all directions of the plane. 
The behavior of the system is determined by the parameters of Hamiltonian (\ref{GBH}), in particular the ratio $U/U_{NN}$ between the 
contact interaction and the strength of the first nearest neighbor (1NN) dipole-dipole interaction. In experiments with Chromium atoms, by using Feshbach resonances, it is possible to control the ratio $U/U_{NN}$ and even turn it down to zero \cite{Lahaye}. In this paper we assume the long range interaction to be a small perturbation with respect to the contact interaction $U/U_{NN} = 20$, because this regime is reachable with Chromium atoms without the need for extreme modification of the scattering length.

\section{Ground state and Metastable states}
\label{sec:GS}
Consider an infinite 2D square lattice reproduced by a $4\times 4$ elementary cell with periodic boundary conditions, filled
with dipolar atoms. Such a system is described by the Hamiltonian (\ref{GBH}) and we know that it is characterized 
by the existence of many almost degenerate metastable states \cite{Menotti}.

Using imaginary time evolution in Eq. (\ref{TDGA}), it is possible to find the ground state of the system. However many times this process gets stucked 
in local minima of energy and in general it is very difficult to reach the actual ground state \cite{Menotti}. This is a clear signature 
of the existence of metastable states. To find all the metastable states, we use a combined mean field and perturbative approach.

We want to write Hamiltonian (\ref{GBH}) as a sum of single-site Hamiltonians. Writing the annihilation operator as $a_i = \tilde{a}_i + \varphi_i$, we 
can perform the mean field decoupling on the product
\begin{eqnarray}
a_i^{\dag}a_j &=& \tilde{a}_i^{\dag} \varphi_j +  \tilde{a}_j \varphi_i + \varphi_i\varphi_j +  \tilde{a}_i^{\dag}\tilde{a}_j \nonumber \\
              &\simeq& a_i^{\dag} \varphi_j + a_j \varphi_i - \varphi_i\varphi_j,
\end{eqnarray}
where in the last step we have assumed small fluctuations, characteristic of the Mott or the deep superfluid states, and replaced $\tilde{a}_i^{\dag}\tilde{a}_j \simeq 0$.
In Hamiltonian (\ref{GBH}) we now replace $a_i^{\dag}a_j$ with the expression calculated above, and find the mean field Hamiltonian
\begin{equation}
H_{MF} = H_0 + H_1,
\label{eqn:HMF}
\end{equation}
where 
\begin{eqnarray}
H_0 &=&  \sum_i \left[ -\mu n_i + \frac{U}{2}  \; n_i(n_i-1) +  \sum_{\vec \ell} \frac{U_{\vec \ell}}{2} \bar{n}_{i,{\vec \ell}} \;  n_i \right], \\
H_1 &=& -J \sum_i \left( \bar{\varphi}_i^* a_i +  \bar{\varphi}_i a_i^{\dag}\right),
\label{GBH_MF}
\end{eqnarray}
and we have neglected terms of the order of $\varphi_i^2$. 

\begin{center}
\begin{figure}[h]
\includegraphics[width=0.8\linewidth]{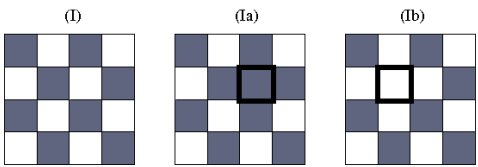}
\caption{Gray sites are occupied by one atom and white sites are empty. A "classical" distribution $|\Phi_I\rangle$ of atoms in the lattice (I). 
The same distribution with one additional atom (Ia) and one removed atom (Ib).}
\label{fig1}
\end{figure}
\end{center}

Given a classical distribution of atoms in the lattice $|\Phi_I\rangle$, e.g. (I) in Fig. \ref{fig1}, that fulfills $H_0 |\Phi_I\rangle = E_0 |\Phi_I\rangle$, 
we want to know whether 
this configuration is stable or not with respect to particle-hole excitations \cite{Fisher}; this holds not only for $|\Phi_I\rangle$ 
being the ground state, but also a metastable state. For each stable configuration there is a region in the $\mu-J$ plane, called Mott lobe, in which the order parameter is zero due to the perfect localization of the atoms at the lattice sites. Therefore, to calculate the Mott lobe of $|\Phi_I\rangle$, we have to evaluate the order parameter 
$\varphi_i = \langle a_i \rangle = \textrm{Tr} (a_i \rho)$ at each site of the lattice. The partition function $Z = \textrm{Tr} (e^{-\beta H_{MF}})$, after a Dyson expansion 
of the exponential, is in the lowest relevant order $Z \simeq \textrm{Tr} (e^{-\beta H_0})$, with $\beta$ being the inverse of temperature. The MF density matrix is $\rho = \frac{1}{Z}e^{-\beta H_{MF}}$,
which in the limit of zero temperature ($\beta \to \infty$) can be expanded around $E_0$, and becomes $\rho = e^{\beta E_0}e^{-\beta H_{MF}}$.

Using again the Dyson expansion of the exponential, we obtain the order parameter as 
\begin{eqnarray}
\varphi_i &\simeq& -e^{\beta E_0} \int_0^\beta \textrm{Tr} \left[a_i e^{-(\beta -\tau)} H_1 e^{-\tau H_0}\right]d\tau = \nonumber\\
	  &=& J \bar{\varphi}_i e^{\beta E_0} \int_0^\beta \textrm{Tr} \left[a_i e^{-(\beta -\tau)}  a_i^{\dag} e^{-\tau H_0}\right]d\tau.
\label{PHIINTEGRAL}
\end{eqnarray}
Performing the integral (\ref{PHIINTEGRAL}) in the zero temperature limit, we trace around $|\Phi_I\rangle$. To avoid the divergence of the integral one has to require that by adding (removing) one particle to (from) $|\Phi_I\rangle$ at any site $i$, as shown in the Fig. \ref{fig1}, the energy increases, i.e. that the state $|\Phi_I\rangle$ is a minimum with respect to particle-hole excitations
in some range of the parameters $J$ and $\mu$.

After simple algebra, one finds the order parameter to fulfill
\begin{eqnarray}
\varphi_i=J \bar{\varphi_i} 
\left[\frac{n_i+1}{Un_i-\mu+V^{1,i}_{dip}}-
\frac{n_i}{U(n_i-1)-\mu+V^{1,i}_{dip}}\right],
\label{pert}
\end{eqnarray}
where $V^{1,i}_{dip}$ is the dipole-dipole interaction of one atom placed at site $i$ with the rest of the lattice, and the 
conditions for convergence are
\begin{equation}
(n_i - 1)U  + V^{1,i}_{dip} \le \mu \le n_i + V^{1,i}_{dip}.
\label{CONVERGENCE}
\end{equation}
One finds such an equation (\ref{pert}), and conditions (\ref{CONVERGENCE}) for every site $i$ of the lattice. The convergence conditions
are simple and among them one has to choose the most stringent to find the boundary of the lobe at $J = 0$. Instead the equations for the order parameters are coupled due to the $\bar{\varphi_i}$ term, which can be written 
in a matrix form $M(\mu,U,J) \cdot \vec{\varphi} = 0$, with $\vec{\varphi} \equiv \left(\cdots\varphi_i\cdots\right)$, and have 
a non trivial solution.
For every $\mu$, the smallest $J$ for which $\textrm{det} \left[M(\mu,U,J)\right] = 0$ gives the lobe of configuration $|\Phi\rangle_I$ in the $\mu-J$ plane. Notice that if the chosen configuration is not stable, one finds that conditions (\ref{CONVERGENCE}) are never satisfied. This is because the requirement that by adding (removing) one particle to (from) $|\Phi_I\rangle$ at any site $i$ the energy increases is false, and the integral (\ref{PHIINTEGRAL}) indeed diverges. 

\begin{center}
\begin{figure}[h]
\includegraphics[width=1.0\linewidth]{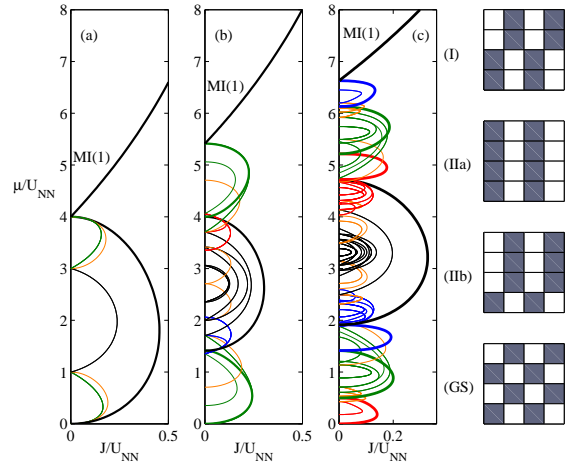}
\caption{(Color online) (a,b,c) Phase diagram with a range of the dipole-dipole interaction cut at the first, second and fourth nearest neighbor respectively. The thick line is
the ground state and the other lobes correspond to the metastable states, the same color corresponding to the same filling factor. In (c) filling factors range from $\nu = 1/8$ to $\nu = 1$. Metastable configuration appearing at the first nearest neighbor (I), and second (IIa-IIb), and the corresponding ground state (GS); the metastable states remain stable for all larger ranges of the dipole-dipole interaction.}
\label{fig:LOBESENERGY}
\end{figure}
\end{center}

We follow the same procedure for every possible classical distribution of atoms in the lattice, for filling factors $\nu = N_a/N_s$
(number of atoms per number of sites) ranging from $\nu = 1/16$ to $\nu = 1$. 
Figs. \ref{fig:LOBESENERGY} (a, to, c) show the phase diagram calculated in this way for a range of the dipole-dipole interaction cut at the first (1NN), second (2NN) and fourth (4NN) nearest neighbor (see also \cite{Goral}).
For 4NN, shown in Fig. \ref{fig:LOBESENERGY} (c), the low tunneling region of the phase diagram consists of many Mott insulating states with different filling factors, ranging from $\nu = 1/8$ to $\nu = 1$. The ground
state (thick line) is a multiple of even filling factors (even number of atoms in the lattice), while metastable states (thin line) show up also with odd values of filling factors apart from $1/16$ and $15/16$. The phase diagram presents almost perfect particle hole duality induced by the strong value of the on-site interaction, meaning that configurations at filling factors bigger than $\nu = 1/2$ show up with the same number and the same distribution of holes in the lattice as the number
and distribution of atoms for filling factors smaller than $\nu = 1/2$. Some of the metastable configurations at $\nu = 1/2$ are presented in Fig. \ref{fig:LOBESENERGY} (I,IIa,IIb) with the 
corresponding ground state (GS).
We also find supersolid domains in the superfluid region of the phase diagram, but we do not consider this issue here.

\subsection{Range of interaction}

When the dipole-dipole interaction is absent ($U_{NN} = 0$), the phase diagram in the low tunneling region, is given  by Mott insulator lobes $MI(n)$ with exactly $n$ particles per site depending on the value of the chemical potential $\mu$ \cite{Fisher}. 
In Fig. \ref{fig:LOBESENERGY} we have shown the phase diagram for a range of dipole-dipole interaction that is cut at the first (1NN) (a) and second (2NN) (b) nearest neighbor. 
Notice that as the range of interaction increases, the lower point of the $MI(1)$ lobe at $J=0$ moves up and lobes for fractional filling 
factors appear in the lower part of the phase diagram. The long range interaction is responsible for the appearance of the lobes under the $MI(1)$ lobe.
The checkerboard starts to appear already at 1NN with a small number of metastable
insulating lobes and if the range of interaction increases the checkerboard moves up in the phase diagram as shown in Fig. \ref{fig:LOBESENERGY}(b). 

As the range of interaction increases new fractional filling factors appear.
For instance in a $4\times 4$ elementary cell the smallest allowed filling factor is $1/16$. By cutting the long range interaction at 4NN we observe that filling factor $1/16$ is not present, because following conditions (\ref{CONVERGENCE}), the configuration of one atom in the $4\times 4$ elementary cell is not stable with respect to particle-hole excitation. Nevertheless this configuration becomes stable for larger ranges of the dipole-dipole interaction.
\begin{center}
\begin{figure}[t]
\includegraphics[width=0.9\linewidth]{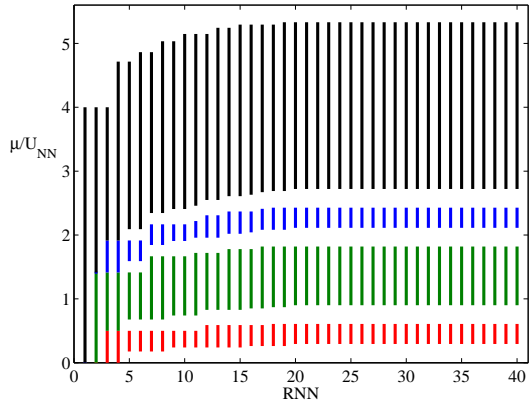}
\caption{(Color online) Boundaries of the GS Mott lobes at zero tunneling, calculated for even filling factors from $2/16$ to $1/2$, as a
function of the range (RNN) of the dipole-dipole interaction. The colors are the same as for the previous figures of the lobes. Notice the discontinuity in the GS after $RNN = 4$ that will be filled by other fractional filling factors.}
\label{fig:BOUNDARIES}
\end{figure}
\end{center}
In Fig. \ref{fig:BOUNDARIES} we plot the boundaries at zero tunneling of the ground state insulating lobes with filling factors multiple of $2/16$, from $2/16$ to $1/2$ as a function of the range of the interaction. Notice that for $RNN \leq 4$ the ground state covers entirely the $\mu$ domain, from $\mu = 0$ up to the maximum value of filling factor $1/2$ zero tunneling boundary, while for larger values of the range of interaction discontinuities 
start to appear; these are filled by other fractional filling factors. 
Notice also that the boundaries at zero tunneling stabilize to steady values when the range of dipole-dipole interaction is sufficiently large.

\subsection{Size of the elementary cell}
The size of the elementary cell also plays an important role in the allowed filling factors. Indeed, in a $N\times N$ cell it is impossible to see filling factors smaller than $1/N^2$. In order to see the ground state of filling factor $1/2$, a $2\times 2$ elementary cell is sufficient, but
 to be able to see filling factors close to zero and, as a consequence of the 
particle hole duality, close to 1, one has to increase the size of the elementary cell. 

\begin{center}
\begin{figure}[h]
\includegraphics[width=0.7\linewidth]{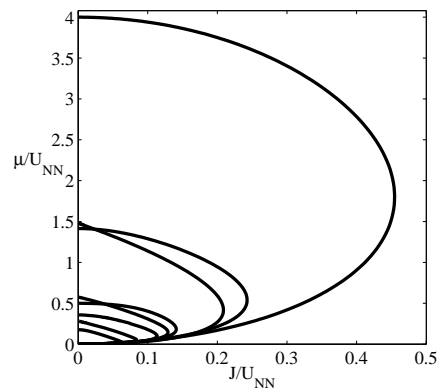}
\caption{Lobes of the GS minimal filling factor for $RNN = 1,...,8$. As the range of dipole-dipole interaction increases the tip of the corresponding lobe gets smaller.}
\label{fig:MATRIOSKA}
\end{figure}
\end{center}

Given a range of the dipole-dipole interaction, there is a rule of thumb to find which is the smallest GS filling factor allowed. It consists of placing atoms in an infinite lattice at the smallest possible interatomic distance compatible with zero dipole-dipole interaction in the system, and find the 
dimension of the elementary cell compatible with this atomic distribution. Table \ref{tab:RANGE} shows the relation between the cut-off range of dipole-dipole interaction $RNN$ and
the GS minimal filling factor $\nu_{GS}$, for a cut-off range of interaction up to the eighth nearest neighbor.
The corresponding lobes in the $\mu-J$ plane are shown in Fig. \ref{fig:MATRIOSKA}.
\begin{center}
\begin{table}[h]
\begin{tabular}{c||ccccccccc}
RNN   & 0 & 1 & 2 & 3 & 4 & 5 & 6 & 7 & 8 \\
\hline
$\nu_{GS}$ & 1 & $\frac{1}{2}$ & $\frac{1}{4}$ & $\frac{1}{5}$ & $\frac{1}{8}$ & $\frac{1}{9}$ & $\frac{1}{10}$ & $\frac{1}{13}$ & $\frac{1}{16}$ 
\end{tabular}
\caption{Range of the dipole-dipole interaction RNN and its corresponding GS minimal filling factor $\nu_{GS}$.}
\label{tab:RANGE}
\end{table}
\end{center}
Notice that as the cut-off range of the long range interaction increases, $\mu_{max}$ shows a tendency to decrease as well as the tip of the lobe for the corresponding minimal filling factor.

In the following, unless differently specified, we will consider the elementary cell to be $4\times 4$ with periodic boundary conditions and the range of dipole-dipole interaction cut 
at the fourth nearest neighbor to cover a sufficiently round region of interaction in the lattice.

\subsection{Non-uniform lattices}
Another interesting thing is to see what happens to an insulating lobe when we add to the two dimensional lattice 
a superlattice, mimicked by a local chemical potential $\Delta\mu_i$ with a specific pattern. This can be useful for applications
such as initialization and manipulation of the metastable states. 

We replace the chemical potential in Eqs. (\ref{pert}-\ref{CONVERGENCE}) with $\mu \rightarrow \mu - \Delta\mu_i$. 
Our convention is that $\Delta\mu_i < 0$ for a deeper well, such that it is energetically favorable for an atom to stay in it. 
For any choice of the $\Delta\mu_i$, one can easily calculate the effect of the superlattice to a given insulating lobe.

\begin{center}
\begin{figure}[h]
\includegraphics[width=1.0\linewidth]{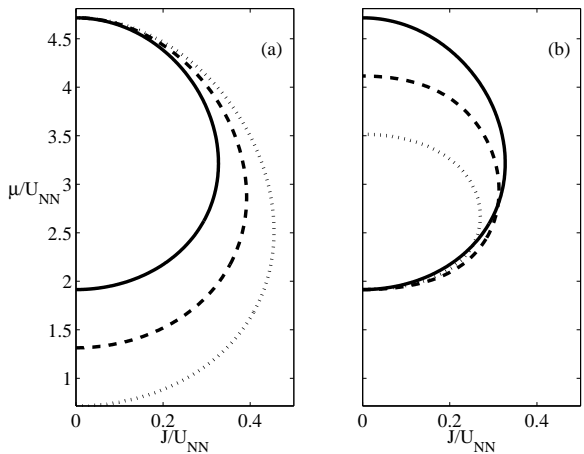}
\caption{Insulating lobes of the checkerboard state with a negative local chemical potential that follow a checkerboard pattern (a), and a stripe pattern (b), see 
text for details. The continuous line shows the lobe without any superlattice applied.}
\label{fig:LOCAL}
\end{figure}
\end{center}

In Fig. \ref{fig:LOCAL}, we show the effect of two different superlattices on the checkerboard insulating lobe.
In Fig. \ref{fig:LOCAL} (a) $\Delta\mu_i = \Delta\mu\;n_i^{(CB)}$, with $\Delta\mu<0$ and $n_i^{(CB)}$ the density of the
checkerboard. The thick, dashed, and dotted lines are for $\Delta\mu/U_{NN} = 0, -0.6, -1.2$ respectively.
As $\Delta\mu$ grows in magnitude,
the lobe becomes bigger as expected. Notice that the upper point of the lobes at $J=0$, does not change while changing $\Delta\mu$, while the lower point moves towards
$\mu = 0$ as $\Delta\mu$ decreases. This is easily understood by looking at inequalities (\ref{CONVERGENCE}), that in this case become
\begin{equation}
(n_i - 1)U  + V^{1,i}_{dip} + \Delta\mu_i \le \mu \le n_i + V^{1,i}_{dip} + \Delta\mu_i.
\label{CONVERGENCE_e}
\end{equation}
The upper limit is given by the smallest of the right hand side (r.h.s.) in conditions (\ref{CONVERGENCE_e}), i.e. at $n_i=0$ where $\Delta\mu_i=0$, while for the lowest limit, we have to choose the biggest of the left hand side
(l.h.s.) condition, where $n_i=1$ and $\Delta\mu_i < 0$.

In Fig. \ref{fig:LOCAL} (b), the local chemical potential follows a stripe pattern $\Delta\mu_i = \Delta\mu\; n_i^{(S)}$, where $n_i^{(S)}$ is the density distribution
of the stripe (S) state (IIa) of Fig. \ref{fig:LOBESENERGY}. The magnitude of $\Delta\mu$ is the same as in the above case for the thick, dashed and dotted lines. 
As $\Delta\mu$ decreases, the lobe becomes smaller due to the distribution of local potential energies that do not favor the checkerboard lobe.
It is not difficult to see that the lower limit, set by the biggest of the l.h.s. of conditions (\ref{CONVERGENCE_e}), is given for $n_i=0$ and $\Delta\mu_i = 0$, 
while the upper limit set by the smallest of the r.h.s. of conditions (\ref{CONVERGENCE_e}), is found where $n_i=1$ and $\Delta\mu_i < 0$.

\subsection{3D lattices}
Optical lattices in real experiments are in general three dimensional and one should take into account that atoms can tunnel in all directions as well as the anisotropic
dipole-dipole interaction with the whole lattice. While it is experimentally feasible to isolate two dimensional layers (2D) such that atoms do not tunnel from one layer 
to the neighboring ones, it
is not possible to switch off the infra-layer dipole-dipole interaction due to its long-range character. However, if the direction of the dipoles is perpendicular to the 
plain of the layers (as in our model), the resulting dipole-dipole interaction between different plains is attractive. In the Mott phase this makes energetically favorable
to have the same distribution of atoms for all layers \cite{Demler}. It would not be difficult to check it by using Eq. (\ref{TDGA}) for a three dimensional system and make use
of the imaginary time evolution technique. This is not the purpose of this work and will be done elsewhere.

\section{Low energy excitations}
\label{sec:EXCITATIONS}
The low-lying excitations are creating particles (p) and holes (h) in a given metastable configuration. For every site $i$, at $J=0$ the 
excitations are given by $E_i^p = Un_i  - \mu + V_{dip}^{1,i}$ and $E_i^h = \mu - U(n_i-1) - V_{dip}^{1,i}$, where $n_i$ is the density at site $i$. Clearly the hole excitation for $n_i=0$ is unphysical. At finite $J$, the excitation spectrum $\omega({\bf k})$ of a metastable configuration, is given by the small fluctuations $\delta f_n^{(i)}(t)$ around the unperturbed metastable state coefficients $\bar{f}_n^{(i)}$. In a Mott state with 
exactly $m_i$ particles at site $i$, the only non-zero coefficients are given by $\bar{f}_m^{(i)}$. Writing 
$f_n^{(i)} = \bar{f}_n^{(i)} + \delta f_n^{(i)}(t)$ in Eq. (\ref{TDGA}), and taking into account only linear terms in the 
fluctuations, we get
\begin{eqnarray}
\label{eqn:FLUCTUATIONS}
i\, \dot{\delta f_n^{(i)}} &\simeq& -J \left[ \bar{\varphi}_i \sqrt{n_i}
  \bar{f}_{n-1}^{(i)} + \bar{\varphi}_i^* \sqrt{n_i+1} \bar{f}_{n+1}^{(i)} \right] +   \\ 
&+& \left[\frac{U}{2} n_i(n_i-1) + n_i V_{dip}^{1,i} -\mu n_i - \chi_{m}^{(i)}\right] \delta f_n^{(i)}, \nonumber
\end{eqnarray}
where $\bar{\varphi}_i \simeq \sum_{\langle j \rangle_i} \sum_n \sqrt{n_j+1} \left(\bar{f}_{n}^{(j)*}\delta f_{n+1}^{(j)} + \bar{f}_{n+1}^{(j)}\delta f_{n}^{(j)*}\right)$, and $\chi_{m}^{(i)} = \frac{U}{2} m_i(m_i-1) + m_i V_{dip}^{1,i} -\mu m_i$ is 
an extra phase that we have introduced to eliminate the rotating phase of the $\bar{f}_m^{(i)}$ coefficients. The only non-trivial
terms in Eq. (\ref{eqn:FLUCTUATIONS}) are therefore
\begin{equation}
\label{eqn:NTFLUCTUATIONS}
\begin{split}
 i\, \dot{\delta f_{m-1}^{(i)}} &= E_i^h \delta f_{m-1}^{(i)} - J\sqrt{m_i}\bar{\varphi}_i^* \\
 i\, \dot{\delta f_{m+1}^{(i)}} &= E_p^h \delta f_{m+1}^{(i)} - J\sqrt{m_i+1}\bar{\varphi}_i, 
\end{split}
\end{equation}
and their complex conjugates. It is convenient to study Eq. \ref{eqn:NTFLUCTUATIONS} and their complex conjugates in the 
Fourier domain with $\delta f_{n}^{(i)}(t) = \sum_k e^{i{\bf k}\cdot {\bf x}^{(i)}} a_n^{(i)}({\bf k},t)$, ${\bf x}^{(i)}$ being the 
2D vector pointing at site $i$.
After simple algebra one finds the Fourier modes to fulfill
\begin{eqnarray}
 i\, \dot{a}_{m-1}^{(i)}({\bf k},t) &=& E_i^h a_{m-1}^{(i)}({\bf k},t) +  \nonumber \\
   &-& J\sum_{\langle j \rangle_i} \left[\sqrt{m_i(m_j+1)}a_{m+1}^{(j)*}({\bf -k},t) + \right.\nonumber \\ 
   &+& \left.\sqrt{m_im_j}a_{m-1}^{(j)}({\bf k},t)\right]e^{i{\bf k}\cdot {\bf d}^{\langle j \rangle}} \label{eqn:FMODES1}\\
 i\, \dot{a}_{m+1}^{(i)}({\bf k},t) &=& E_i^p a_{m+1}^{(i)}({\bf k},t) +  \nonumber \\
   &-& J\sum_{\langle j \rangle_i} \left[\sqrt{(m_i+1)(m_j+1)}a_{m+1}^{(j)}({\bf k},t) \right. + \nonumber \\
   &+& \left. \sqrt{(m_i+1)m_j}a_{m-1}^{(j)*}(-{\bf k},t)\right] e^{i{\bf k}\cdot {\bf d}^{\langle j \rangle}} \label{eqn:FMODES2},
\end{eqnarray}
with ${\bf d}^{\langle j \rangle} = \left\{\pm(d,0),\pm(0,d)\right\}$ being the vector of nearest neighbors in the lattice,
and $d$ is the lattice spacing. We look for stationary solutions of Eqs. (\ref{eqn:FMODES1},\ref{eqn:FMODES2}) with the ansatz
$a_{n}^{(i)}({\bf k},t) = u_{n}^{(i)}({\bf k})e^{-i\omega({\bf k})t} + v_{n}^{(i)}({\bf k})e^{i\omega({\bf k})t}$.
For every site $i$ of the elementary cell, Eqs. (\ref{eqn:FMODES1},\ref{eqn:FMODES2}) become
\begin{widetext}
\begin{equation}
\label{eqn:COUPLED01}
\left\{ \begin{array}{l}
 \left[E_i^h - \omega({\bf k})\right]u_{m-1}^{(i)}({\bf k}) - J \sum_{\langle j \rangle_i} \left[\sqrt{m_i(m_j+1)}v_{m+1}^{(j)*}({\bf -k}) + 
        \sqrt{m_im_j}u_{m-1}^{(j)}({\bf k})\right]e^{i{\bf k}\cdot {\bf d}^{\langle j \rangle}} = 0 \\
\left[E_i^p + \omega({\bf k})\right]v_{m+1}^{(i)*}({\bf k}) - J \sum_{\langle j \rangle_i} \left[\sqrt{(m_i+1)(m_j+1)}v_{m+1}^{(j)*}({\bf -k}) + \sqrt{(m_i+1)m_j}u_{m-1}^{(j)}({\bf k})\right]e^{i{\bf k}\cdot {\bf d}^{\langle j \rangle}} = 0\\
 \left[E_i^p - \omega({\bf k})\right]u_{m+1}^{(i)}({\bf k}) - J \sum_{\langle j \rangle_i} \left[\sqrt{(m_i+1)(m_j+1)}u_{m+1}^{(j)}({\bf k}) + 
        \sqrt{(m_i+1)m_j}v_{m-1}^{(j)*}({\bf -k})\right]e^{i{\bf k}\cdot {\bf d}^{\langle j \rangle}} = 0 \\
\left[E_i^h + \omega({\bf k})\right]v_{m-1}^{(i)*}({\bf -k}) - J \sum_{\langle j \rangle_i} \left[\sqrt{m_i(m_j+1)}u_{m+1}^{(j)}({\bf k}) + \sqrt{m_im_j}v_{m-1}^{(j)*}({\bf -k})\right]e^{i{\bf k}\cdot {\bf d}^{\langle j \rangle}} = 0.
\end{array} \right.
\end{equation}
\end{widetext}
This set of $4N^2$ equations can be reduced depending on the symmetry of the density distribution, like in the case of the checkerboard where only two sites are important. Eqs. (\ref{eqn:COUPLED01}) can be written in a matrix form, 
$M \left( \begin{array}{l}
{\bf u} \\
{\bf v^*}
\end{array}
\right) = 0$, and have non-trivial solution 
only if $\textrm{det}\left[M\right] = 0$. The excitation spectrum is then given by the positive solutions of the last equation. We have checked that Eqs. (\ref{eqn:COUPLED01}) lead to an excitation spectrum that perfectly agrees with the one calculated in \cite{sinha} for the checkerboard and the $MI(n)$ states. 
In Fig. \ref{fig:POPULATION} (a), we show the
lowest excitation branch of the four metastable configurations of Fig. \ref{fig:LOBESENERGY}, for $\mu = 3.3U_{NN}$, $J = 0.1U_{NN}$ and $k_xd = k_yd = k/\pi$, in the first Brillouin zone.
The thick line is for the (CB) state, the dashed, dash-dotted and dotted lines are for (I), (IIa) and (IIb) states respectively.
At the boundaries of the insulating lobes the excitation spectrum $\omega({\bf k} = 0)$ goes to zero.

\subsection{Oscillations}
In the real time evolution in Eq. (\ref{TDGA}), at a constant density, the chemical potential $\mu$ gives only a phase factor,  therefore the tunneling coefficient
$J$ is the only important parameter. Suppose at time $t$ the system is described by a Gutzwiller state $\ket{\phi_{t}}$, we define the population of the metastable state $\ket{\phi_{MS}}$ as 
\begin{equation}
\label{eqn:POPULATION}
P_{MS}(t) = \sqrt[N_s]{|\langle\phi_{MS}\ket{\phi_{t}}|^2},
\end{equation}
the $N_s^{th}$ root of the fidelity, where $N_s$ is the number of sites of the elementary cell. This definition has the advantage on the fidelity that it does not depend on the number of sites, while the simple fidelity would be one if and only if $\ket{\phi_{t}} = \ket{\phi_{MS}}$, and otherwise depend on the dimension of the cell and tend to zero for an infinite number of sites $N_s$.

\begin{center}
\begin{figure}[h]
\includegraphics[width=0.9\linewidth]{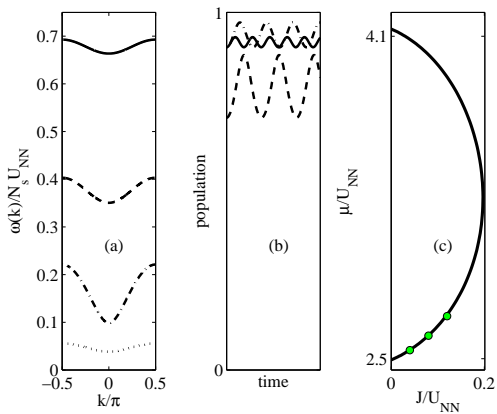}
\caption{(a) Lowest excitation spectrum of metastable state (GS) (thick), (I) (dashed), (IIa) (dash-dotted) and (IIb) (dotted) of 
Fig. \ref{fig:LOBESENERGY} calculated for $\mu = 3.3U_{NN}$, and $J = 0.1U_{NN}$. Population (b) of the metastable state (I) during real time evolution, the thick and dash-dotted lines are for a small perturbation of
the metastable state for $J = 0.04U_{NN}$ and $J = 0.12U_{NN}$, while the dashed line corresponds to an initial big perturbation and 
$J = 0.08U_{NN}$; (c) the Mott insulating lobe of the state (I).}
\label{fig:POPULATION}
\end{figure}
\end{center}

In a metastable state, atoms are perfectly localized at the sites of the lattice, and the system is Mott insulator. By adding some
''noise'' in the GW coefficients we randomly remove population from occupied sites and move it to empty ones, conserving the total number of atoms, and the system is superfluid. 
In general, for a given initial condition close to a metastable state (meaning that the density follows the distribution of the metastable state plus some noise)
and the tunneling coefficient $J$ smaller than the tip $J_{tip}$ of
the metastable insulating lobe, in the real time evolution 
we observe small oscillations around a local minimum of the energy with a multi-component frequency $\upsilon$. The frequency of 
oscillation $\upsilon$ depends on the exact initial condition and on the tunneling coefficient. As an example in Fig. \ref{fig:POPULATION} (b) we show the real time dynamics of the population of metastable state (I). 
 The thick and dash-dotted lines
are calculated for $J = 0.04U_{NN}$ and $J = 0.12U_{NN}$ respectively, the approximate 
oscillation frequencies, in units of $\hbar = 1$, are given by $\upsilon \simeq 2\pi/150$ and $\upsilon \simeq 2\pi/227$ respectively. The dashed line 
is calculated for $J = 0.08U_{NN}$ and at a larger value of the initial perturbation, so that its oscillation frequency is $\upsilon \simeq 2\pi/297$.
In Fig. \ref{fig:POPULATION} (c) we plot the insulating lobe of configuration (I), and the round spots are placed in correspondence of the values of the parameters for oscillations shown in Fig.~(\ref{fig:POPULATION}) (b).

\section{Stability of the metastable states}
\label{sec:STABILITY}
In Sect. (\ref{sec:GS}) we have studied an infinite 2D lattice reproduced by a $4\times 4$ elementary cell with periodic boundary conditions. We have shown that polarized dipolar
bosons in such a system feature many almost degenerate metastable states that are stable against particle-hole excitations. Another clear sign of the existence of metastable
configurations is the fact that in the imaginary time evolution is very difficult to reach the ground state and often the process gets stucked in local minima of energy.
Therefore, we can think of these states as local minima of a potential where a particle can be trapped for a 
certain time that depends on the barrier that separates it from another local minimum.
 
Before studying the stability of the metastable states we remind the simple case of a particle in a double well potential.
\begin{center}
\begin{figure}[h]
\includegraphics[width=0.8\linewidth]{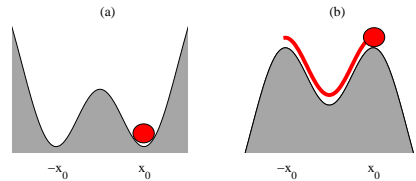}
\caption{Particle in a double well (a), and the instanton (b).}
\label{fig:INSTANTON}
\end{figure}
\end{center}
Being the particle at time $t=0$ in the right well ($x_0$) as shown in Fig. \ref{fig:INSTANTON} (a), the probability 
at time $T$ for the particle to tunnel in the left well can be calculated using the propagator in imaginary time and a path integral approach \cite{Wen}. The probability amplitude 
for the particle to tunnel is given by
\begin{equation}
\langle -x_0 |e^{-TH} |x_0 \rangle = \sinh (T\omega_0 e^{-S_0}),
\label{INSTANTON}
\end{equation}
where $\omega_0$ is of the order of the frequency at which the particle oscillates around the local minimum $x_0$, and $S_0$ is the minimal action along the stationary 
path that connects $x_0$ to $-x_0$ in the inverted potential of Fig. \ref{fig:INSTANTON} \index{}(b), called an instanton.

The corresponding probability amplitude in real time is obtained from (\ref{INSTANTON}) by analytical continuation just by replacing $T = iT$, and one finds that the particle has tunneled completely to the left well after a time given by
\begin{equation}
 T\omega_0 = \frac{\pi}{2}e^{S_0}.
\end{equation}  
Therefore the tunneling time is known once we know the frequency of small oscillations $\omega_0$ and the action $S_0$ along the stationary path in the inverted potential.

\begin{center}
\begin{figure}[h]
\includegraphics[width=0.7\linewidth]{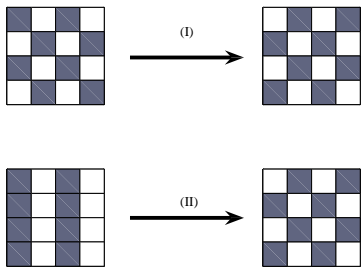}
\caption{(I) Exchanging particles with holes, and (II) process where only in a region of the lattice (first and third row from top) the exchange of particles with holes takes place.}
\label{fig:TUNNELING}
\end{figure}
\end{center}

Regarding the metastable states, the analogy of the tunneling of a particle in the double well potential is the process in which a metastable state tunnels into its complementary, in which the role of particles and holes is exchanged as shown in Fig. \ref{fig:TUNNELING} (I). 
Given a metastable configuration defined by its Gutzwiller coefficients $\{f_n^{(i)}\}$, it is not straightforward to identify the barrier that separates it from its complementary.  For the Gutzwiller wavefunction we look for a simple parametrization that allow us to identify the metastable states and parametrize in a simple way the process of exchanging atoms with holes and viceversa in certain lattice sites. 
We will describe the process of passing from one local minimum to another one using only one variable and its conjugate momentum.
The process in which a metastable state tunnels into a state different from its complementary, as shown
in Fig. \ref{fig:TUNNELING} (II), carries the complication that initial and final states are not degenerate.

\subsection{Parametrization and Ansatz}
The imaginary time Lagrangian of a system \cite{Perez}, described by a quantum state $|\Phi\rangle$,  
is given by
\begin{equation}
\mathcal{L} = - \frac{\langle\dot{\Phi}|\Phi\rangle - \langle\Phi|\dot{\Phi}\rangle}{2} + \langle\Phi|H|\Phi\rangle.
\end{equation}
The coefficients of the Gutzwiller wavefunction (\ref{GW}) in general can be complex numbers. We write them in this way
\begin{equation}
\begin{split}
f_n^{(i)}  &= \frac{1}{\sqrt{2}}\left(x_n^{(i)} + ip_n^{(i)}\right),\\
f_n^{*(i)} &= \frac{1}{\sqrt{2}}\left(x_n^{(i)} - ip_n^{(i)}\right),
\end{split}
\end{equation}
where $x_n^{(i)}$ and $p_n^{(i)}$ are real numbers. With the last prescription, the Lagrangian of the system becomes
\begin{equation}
\mathcal{L}(x_n^{(i)},p_n^{(i)}) = -i\sum_{i,n=0}^1 p_n^{(i)}\dot{x}_n^{(i)} + \langle\Phi|H(x_n^{(i)},p_n^{(i)})|\Phi\rangle .
\end{equation}
We write it in its canonical form
\begin{equation}
\mathcal{L}(x_n^{(i)},P_n^{(i)}) = \sum_{i,n=0}^1 P_n^{(i)}\dot{x}_n^{(i)} - \mathcal{H}(x_n^{(i)},P_n^{(i)}),
\label{LAGRANGIAN:TOT}
\end{equation}
where
\begin{equation}
\begin{split}
P_n^{(i)} &= -ip_n^{(i)}, \quad \text{and} \\
\mathcal{H}(x_n^{(i)},P_n^{(i)}) &= -\langle\Phi|H(x_n^{(i)},p_n^{(i)})|\Phi\rangle
\end{split}
\end{equation}
is the conserved quantity.

Lagrangian (\ref{LAGRANGIAN:TOT}) is an equation in $2N_S$ independent variables and their conjugate momenta , where $N_S$ is the number of sites of the lattice. We want to reduce the number of independent variables to one.

Consider a simpler case when we have only two sites with one particle in the left well (O=occupied) and we want to parametrize the process in which the particle
tunnels into the right well (E=empty), with the constraints on $(x_n^{(i)},p_n^{(i)})$ that the normalization is respected and the number of atoms is conserved
\begin{equation}
\begin{split}
\sum_{n=0}^1 \frac{1}{2}\left({x_n^{(i)}}^2 - {P_n^{(i)}}^2\right) &= 1,  \qquad i = O,E\\
\sum_{i = O}^E \frac{1}{2}\left({x_1^{(i)}}^2 - {P_1^{(i)}}^2\right) &= 1.
\end{split}
\label{CONDITIONS}
\end{equation}
We make the following ansatz
\begin{equation}
\begin{split}
x_1^{(E)} &= x_0^{(O)} = q,\\
P_1^{(E)} &= P_0^{(O)} = P,\\
P_0^{(E)} &= P_1^{(O)},\\
x_0^{(E)} &= x_1^{(O)},\\
P_1^{(O)} &= -P_0^{(O)}.
\end{split}
\label{ANSATZ}
\end{equation}
The only independent variable is $q$. Its conjugate momentum $P$ is a complicated function given by $\partial\mathcal{L}/\partial\dot{q}$.
When $(q,P) = (0,0)$ the atom is in the left well (O) while at $(q,P) = (\sqrt{2},0)$ it is in the right one (E). The instanton is then the stationary path that joins those two points in phase space. The action $S_0$ is calculated along this path.

For more complicated cases, as the processes described by Figs. \ref{fig:TUNNELING}, there are more than two sites that exchange particles with holes and 
viceversa. We consider only two independent sites (O) and (E), subject to the parametrization $(q,P)$ explained above, where (O) is occupied by one atom and in (E) there is a hole. The remaining occupied sites ($j \in \{O\}$), behave as the independent (O), while the empty ones ($j \in \{E\}$) behave as the independent (E). We also have to take into account potential
sites that do not change, as in the example of Fig. \ref{fig:TUNNELING} (II).
These conditions together with ansatz (\ref{CONDITIONS}) and (\ref{ANSATZ}), enter in in Eq. (\ref{LAGRANGIAN:TOT}) as Lagrange multipliers, and the Hamiltonian becomes
\begin{widetext}
\begin{eqnarray}
 \mathcal{H} (q,P) = \mathcal{H} (q,P) + \lambda_1 \left[\sum_i\sum_{n=0}^1 n \left({x_n^{(i)}}^2 - {P_n^{(i)}}^2\right) - 2\right] + \lambda_2 \left(P_1^{(O)}+P\right) 
+ \lambda_3\left(x_0^{(E)} - x_1^{(O)}\right) + \nonumber\\
+ \lambda_4 \left(P_0^{(E)}-P_1^{(O)}\right) + \lambda_5 \left(x_1^{(E)}-q\right) + \lambda_6 \left(P_1^{(E)}-P\right) + \nonumber\\
+ \sum_{j\in \{O\}} \left[ \lambda_{j,0}^O\left(x_0^{(j)}-q\right) + \lambda_{j,1}^O\left(x_1^{(j)}-x_1^{(O)}\right)  
+ \eta_{j,0}^O\left(P_0^{(j)} - P\right) + \eta_{j,1}^O\left(P_1^{(j)} -P_1^{(O)}\right)\right] + \nonumber\\
+ \sum_{j\in \{E\}} \left[ \lambda_{j,0}^E\left(x_0^{(j)}-x_0^{(E)}\right) + \lambda_{j,1}^E\left(x_1^{(j)}-q\right)  
+ \eta_{j,0}^E\left(P_0^{(j)} - P_0^{(E)}\right) + \eta_{j,1}^E\left(P_1^{(j)} -P\right)\right] + \nonumber\\
+ \sum_{j\in \{0\}} \left[ \lambda_{j,0}^0\left(x_0^{(j)}-2\right) + \lambda_{j,1}^0x_1^{(j)} + \eta_{j,0}^0P_0^{(j)} + \eta_{j,1}^0P_1^{(j)}\right] + \nonumber\\
+ \sum_{j\in \{1\}} \left[ \lambda_{j,0}^1x_0^{(j)} + \lambda_{j,1}^1\left(x_1^{(j)}-2\right) + \eta_{j,0}^1P_0^{(j)} + \eta_{j,1}^1P_1^{(j)}\right],
\label{MONSTER}
\end{eqnarray}
\end{widetext}
where the first two lines are for the independent sites (O) and (E), the third and fourth correspond to sites forced to behave like (O) or (E), and in the last two lines 
we have taken into account also possible conditions for sites that do not change.

\subsection{Action and barrier}
Given an initial and final distribution of atoms in the lattice, as for the examples explained above, the procedure will be: (i) identify the sites that exchange particles with holes and viceversa, sites that do not change, and adjust Lagrange multipliers in Eq. (\ref{MONSTER}); (ii) calculate the stationary path that starts at $(q,P) = (0,0)$ and, for 
degenerate states, ends at $(q,P) = (\sqrt{2},0)$; (iii) calculate the action along the path, given by
\begin{equation}
S_0 = \int \mathcal{L}(q,P)d\tau = \int_{PATH}\mathcal{L}(q,P)\frac{dq}{\dot{q}},
\end{equation}
with $\dot{q} = \partial \mathcal{H}/\partial P$ from Eq. (\ref{MONSTER}). 

\begin{center}
\begin{figure}[h]
\includegraphics[width=0.9\linewidth]{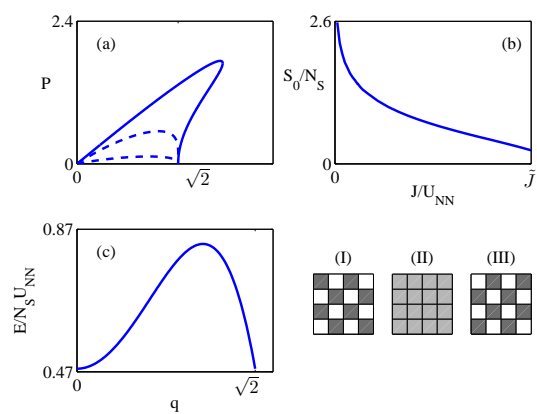}
\caption{Stationary paths (a), action per site (b), and barrier (c). The initial state (I) at $(q,P) = (0,0)$ and final state (III) at $(q,P) = (\sqrt{2},0)$. The state
in (II) is at a middle point $(q,P) = (1,0)$.}
\label{ACTIONCB}
\end{figure}
\end{center}

\begin{center}
\begin{figure}[h]
\includegraphics[width=0.9\linewidth]{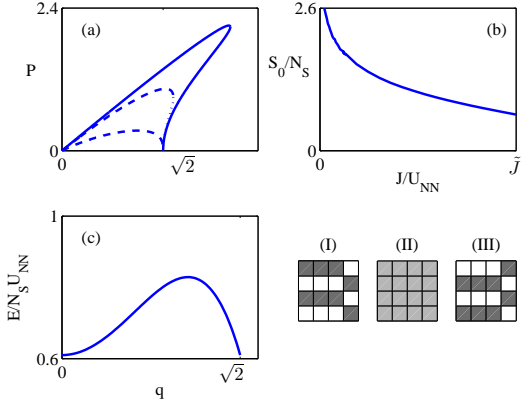}
\caption{Stationary paths (a), action per site (b), and barrier (c). The initial state (I) at $(q,P) = (0,0)$ and final state (III) at $(q,P) = (\sqrt{2},0)$. The state
in (II) is at a middle point $(q,P) = (1,0)$.}
\label{ACTIONL}
\end{figure}
\end{center}

\begin{center}
\begin{figure}[h]
\includegraphics[width=0.9\linewidth]{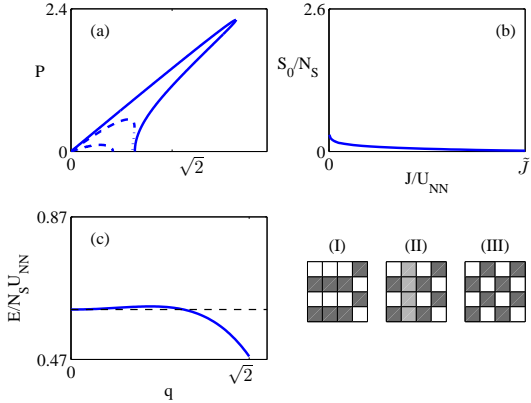}
\caption{Stationary paths (a), action per site (b), and barrier (c). The initial state (I) at $(q,P) = (0,0)$ and final state (III) at $(q,P) = (\sqrt{2},0)$. The state
in (II) is the decay point at $(q,P) \simeq (0.89,0)$.}
\label{ACTIONLCB}
\end{figure}
\end{center}

\begin{center}
\begin{figure}[h]
\includegraphics[width=0.9\linewidth]{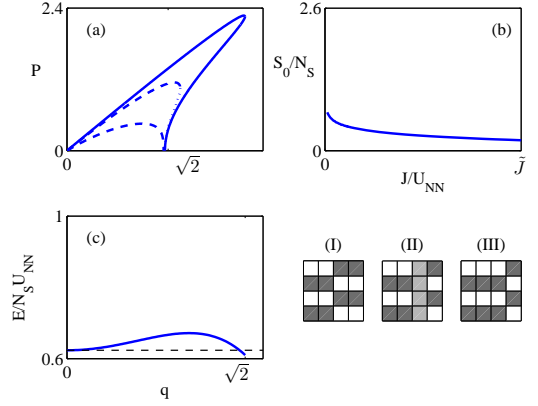}
\caption{Stationary paths (a), action per site (b), and barrier (c). The initial state (I) at $(q,P) = (0,0)$ and final state (III) at $(q,P) = (\sqrt{2},0)$. The state
in (II) is the decay point at $(q,P) \simeq (1.36,0)$.}
\label{ACTIONPL}
\end{figure}
\end{center}

In Figs. \ref{ACTIONCB}, \ref{ACTIONL}, we show the stationary path (a) that connects the metastable state (I) with its complementary (III). At $J \rightarrow 0$ (thick line), the path is bigger than at $J \neq 0$ (dashed line), then the orbit reaches its minimum extension in correspondence of the tip of the lobe $\tilde{J}$. In Figs. \ref{ACTIONCB}, \ref{ACTIONL} (b), we plot the action per site as a function of the tunneling coefficient. The action diverges at $J \rightarrow 0$ and it reaches its minimum in correspondence of the tip of the metastable lobe. 
In Figs. \ref{ACTIONCB}, \ref{ACTIONL} (c), we calculate the barrier as $-\mathcal{H} (q,P=0)$ for $J=0$.
In Figs. \ref{ACTIONLCB}, \ref{ACTIONPL}, we plot the path, the action and the barrier for a metastable state (I) that tunnels into a non-degenerate one (III). Notice that in contrast with the previous case, the paths start at $(q,P) = (0,0)$ but end at $0<q<\sqrt{2}$. This is due to the fact that the final and initial states
do not have the same energy.


By comparing the above figures we observe that given
a metastable state, a longer lifetime corresponds to
a lower energy barrier. Small energy differences between
the initial and the final states and large regions of the lattice
undergoing particle-hole exchange in the tunneling process
contribute to large energy barriers.
Hence, in general it is more likely for a given
state to tunnel into a state deeper in energy, e.g. the
ground state, than into its complementary, which implies
the exchange of particles with holes in the whole lattice.

\section{Dynamics}
\label{sec:DYNAMICS}
Once the lattice is prepared in a configuration with a certain symmetry, the capability of manipulating the configuration is essential in order to use the system as 
a quantum memory. Given an initial metastable state, it would be nice to change in time the lattice parameters such that the system evolves in a deterministic way towards
another chosen metastable configuration. Since there are many metastable states and many parameters, we study the problem in a simplified scenario of a finite $2\times 2$ square lattice, and with a cut-off range of the dipole-dipole interaction at the second nearest neighbor (2NN).

\begin{center}
\begin{figure}[h]
\includegraphics[width=0.7\linewidth]{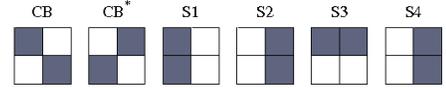}
\caption{Disposition of two atoms in the lattice. In MF, CB and CB$^*$ is the ground state, while S1,..,S4 are degenerate metastable states.}
\label{fig:BASIS}
\end{figure}
\end{center}
In this section, we study how to dynamically pass from a given  configuration to another one, with two different methods: (i) adiabatic passage in which we look for an adiabatic transfer of one state into another one, and (ii) through MF real-time evolution. 

\subsection{Adiabatic passage}
We study the exact Bose-Hubbard Hamiltonian (\ref{GBH}) for the $2\times 2$ lattice mentioned above at filling factor 1/2. There are 6 possible ways of placing two atoms in the lattice and are shown in Fig. \ref{fig:BASIS}, which provide the basis for the Hilbert subspace of the Bose-Hubbard Hamiltonian. Moreover, the MF phase diagram of such a system \cite{FootNote01} consists of a 2-times degenerate checkerboard (CB,CB$^*$) ground state and a 4-times degenerate metastable state in which particles are vertically or horizontally aligned in the lattice as a stripe (S1,..,S4) pattern. The CB lobe and S lobes are shown in Fig. \ref{fig:LOBESENERGY}(b). 

The Hamiltonian in this basis is non diagonal because of tunneling. By diagonalizing the Hamiltonian, we find the ground state
\begin{equation}
\label{eqn:BHGS}
|\psi_{GS} \rangle = x\sum_{i=1}^4 |\psi_{Si} \rangle + \sqrt{\frac{1-4x^2}{2}}\left(|\psi_{CB} \rangle + |\psi_{CB^*} \rangle\right)
\end{equation}
to be a symmetric combination of all the states of the basis, where $x$ is a function of the tunneling coefficient.

Now we want to add to the lattice a superlattice mimicked by a local chemical potential
\begin{equation}
\label{eqn:LOCAL}
\mu - \Delta\mu_i = \mu - \delta\mu \;n_i^{CB} - \Delta\mu \;n_i^{S1},
\end{equation}
where $n_i^{CB}$ and $n_i^{S1}$ are respectively the density distributions of CB and S1 state of Fig. \ref{fig:BASIS}.

\begin{center}
\begin{figure}[h]
\includegraphics[width=1.0\linewidth]{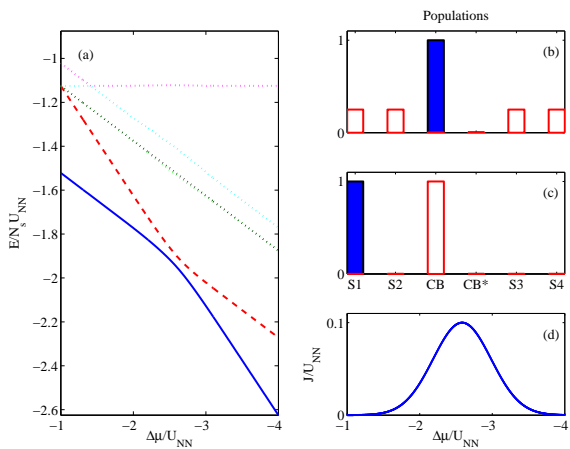}
\caption{(Color online) (a) Eigenenergies of the Bose-Hubbard Hamiltonian, thick and dashed line are the ground and first excited state, as a function of $\Delta\mu$. (b) The ground state of the system is initially prepared in a CB
state (filled bar), while the first excited state consists of equally populated stripe states (empty bars), then an S1-type local chemical potential is applied. (c) The final populations is given by an S1 ground state (filled bar) and a  CB first excited state (empty bar). (d) The variation of the tunneling coefficient.}
\label{fig:ANTICROSSINGCBS}
\end{figure}
\end{center}

To transfer a CB state to the metastable S1, the procedure is: (i) prepare the system in CB; one has to break the symmetry of the ground state 
(\ref{eqn:BHGS}) by applying a local chemical potential that privileges CB, in this case $\delta\mu < 0$ and $\Delta\mu = 0$ in Eq. (\ref{eqn:LOCAL}). Notice
that $\delta\mu$ from now on will be kept constant. 
Then (ii), apply a second local negative chemical potential $\Delta\mu$ in the position of atoms in S1 state. The last process is shown in Fig. \ref{fig:ANTICROSSINGCBS}, where we plot (a) the spectrum of the Bose-Hubbard Hamiltonian as a function of $\Delta\mu$. 
At $\Delta\mu = 0$ the system is already prepared in the CB state for $\delta\mu = -U_{NN}$ as shown from the population graph Fig. \ref{fig:ANTICROSSINGCBS} (b). As $\Delta\mu$ decreases the population is adiabatically
transferred in the S1 state Fig. \ref{fig:ANTICROSSINGCBS} (c). In correspondence to the anticrossing we have increased the tunneling coefficient $J$ as shown in Fig. \ref{fig:ANTICROSSINGCBS} (d), in order to increase the magnitude of the gap. 
For a different adiabatic transfer involving other states, one has to go through the above steps with the appropriate "distribution" of local chemical potential
in the lattice.

\begin{center}
\begin{figure}[h]
\includegraphics[width=1.0\linewidth]{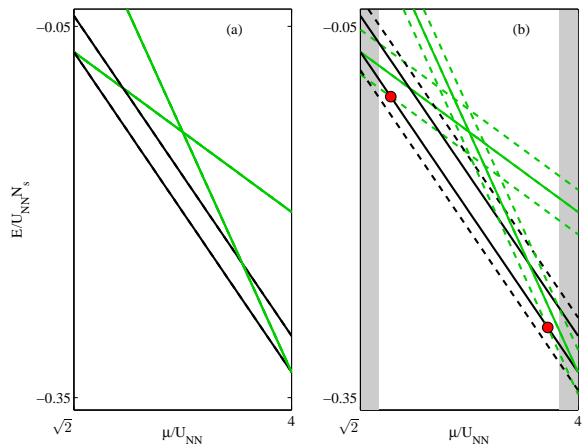}
\caption{Spectrum of the Bose-Hubbard Hamiltonian for $J=0$ (a), and $J = 0.18U_{NN}$ (b). At $J=0$ the ground state coincides with the MF ground state. In (b) the
shaded areas represent the superfluid region calculated in MF, while the round spots are the phase boundaries of filling factor $\nu = 1/2$ calculated with the 
Bose-Hubbard Hamiltonian.}
\label{fig:LOBESBH}
\end{figure}
\end{center}

The drawback of this approach is that for a finite system there is no SF-MI phase transition \cite{Sachdev}, therefore no concept of MI lobes. Nevertheless
we can identify traces of the ground state MI lobe for the filling factor $1/2$. We calculate the spectrum of Hamiltonian (\ref{GBH}) for the $2\times 2$ system introduced
above at $\Delta\mu_i = 0$ but for all filling factors, and in Fig. \ref{fig:LOBESBH} we plot the eigenenergies versus the chemical potential. The different slopes are for the different filling factors $\nu = 1/4,1/2$, and $3/4$ respectively.
At $J=0$ Fig. \ref{fig:LOBESBH} (a) there are 4 degenerate eigenstates both for filling factor $\nu = 1/4$ and $\nu = 3/4$, while for $\nu = 1/2$ the
number of eigenstates is six: a twice degenerate ground state, and an excited state manifold of four degenerate states. When $J$ becomes non-zero (b) the degeneracy
breaks, $\nu = 1/4$ and $\nu = 3/4$ both split into three levels while $\nu = 1/2$ splits into four.
We identify two types of eigenstates: (i) Mott-like states for which the eigenenergies do not depend on the tunneling coefficient $J$ (continuous lines), and (ii) superfluid-
like states that change their eigenenergies as $J$ increases (dashed lines). 
At $J=0$, as shown in Fig. \ref{fig:LOBESBH} (a), the ground state boundaries of $\nu = 1/2$ range from $\mu = \sqrt{2}U_{NN}$ to $\mu = 4U_{NN}$ and coincide with the boundaries of the checkerboard calculated 
in MF. 
At $J\neq 0$, as shown in Fig. \ref{fig:LOBESBH} (b), we estimate the boundaries of the Mott-like state (first thick line) as the crossing points with the superfluid-like ground states of $\nu = 1/4$ and 
$\nu = 3/4$ (dashed lines), plotted as round spots in the graph. The shaded area is the MF superfluid region around filling factor $1/2$. 
The boundaries of the checkerboard calculated in MF, enclosed in the shaded area, are different due to the low accuracy of this method.

\subsection{MF real time evolution}
For large lattices, it is more reliable to look at the dynamics in MF. To pass from one configuration to another, it turns out to be necessary to go into the superfluid region of the phase 
diagram. Even if at the MI-SF transition it is impossible to be adiabatic because of the continuous excitation spectrum of the SF phase \cite{sinha}, for a certain range of lattice parameters the process works.
  
We describe the dynamics through Eq. (\ref{TDGA}) in real time. Two things are worth noticing: (i) to have a non trivial dynamic, one has to prepare an initial state
with a non zero superfluid parameter $\varphi_i \neq 0$, and of course $J \neq 0$. This is because the coupling term in Eq. (\ref{TDGA}) is directly proportional to 
the order parameter. And (ii), since the number of particles is a constant of the motion that is fixed from the initial condition, the chemical potential $\mu$ gives only a phase factor during the evolution. As a consequence of the latter, at a constant integer density the only important point in the phase diagram is the tip $J_{tip}$ of the insulating lobe indicating the phase transition at constant density, such that for $J$ smaller than $J_{tip}$ the system shows small oscillations around a local minimum of the energy, as presented in Fig. \ref{fig:POPULATION}, while for larger values of $J$, one finds deep superfluid oscillations.

With the population defined as in Eq. (\ref{eqn:POPULATION}), we aim to transfer population from a given metastable configuration to another one with a different symmetry, 
by changing the lattice parameters. In the MF regime, the Mott insulator states are exact eigenstates of the MF Hamiltonian. 
The effect of a non uniform lattice on a given configuration, is only to change the size of its insulator lobe but, as long as the lobe exists, the configuration remains stable.
This is consistent with the definition of metastable states as local minima of energy and is confirmed by the lobes of Fig. \ref{fig:LOCAL}. This means that there is no coupling between them and consequently, by changing a local chemical potential 
$\Delta\mu_i$ as in Eq. (\ref{eqn:LOCAL}), there is no possibility of having an anticrossing of the type of Fig. \ref{fig:ANTICROSSINGCBS} but only a perfect crossing. 
The only possibility of transferring population from a metastable state to another one, is by passing through the superfluid region (SF) of the phase diagram and enter a different metastable insulating lobe. It turns out that this is a quantum controlled process which is very much sensitive to the exact initial conditions and the way the parameters change in time.

We specifically study the $2\times 2$ system introduced above with periodic boundary conditions and again, we want to transfer population from the CB state to S1 by applying 
the time dependent local chemical potential $\Delta\mu_i$ of Eq. (\ref{eqn:LOCAL}), in favor of S1. We also want to change the tunneling coefficient $J$ in time, so 
as to exit the CB lobe and, through the superfluid region, enter into the S1 lobe. Ideally we want the population of S1 at the end of the process to be one, 
$P_{S1}(t_{fin}) = 1$, but the actual value of $P_{S1}(t_{fin})$ is very much sensitive on the exact values the parameters take during the dynamics. 
Specifically, we change the lattice parameters smoothly in time as
\begin{equation}
\Delta\mu(t) = -C\tanh\left[\frac{\alpha}{C}\left(t - t_0\right)\right] + C\tanh\left[-\frac{\alpha}{C}t_0\right],
\end{equation}
where $C=1.7 U_{NN}$ and $t_0=60/U_{NN}$ (in units of $\hbar = 1$), are kept constant, while $\alpha$ is a free parameter that sets the maximum slope for this 
function, and
\begin{eqnarray}
&J(\Delta\mu)& = \frac{J_{m}-J_0}{2} \times \\
&\min& \left\{ \begin{array}{l}
\tanh\left[-s m_{o}\right]-\tanh\left[s\left(\Delta\mu - m_{o}\right)\right]  + \frac{2J_0}{J_{m}-J_0} \nonumber\\
\tanh\left[s\left(\Delta\mu - m_{i}\right)\right] - \tanh\left[-s m_{i}\right] + \frac{2J_m}{J_{m}-J_0}, \nonumber
\end{array} \right.
\end{eqnarray}
with $s=15/U_{NN}$ and $J_0 = 0.02U_{NN}$ constants, $m_{i} = -2.6U_{NN}$ fixes the superfluid to Mott insulator transition point at $\Delta\mu_{in} = -2.57U_{NN}$, whereas $J_m$ and
$m_{o}$ are free parameters related with the maximum value of tunneling coefficient in the superfluid region and the point $\Delta\mu_o$ where the CB ceases to exist. Together with the intensity $I_r$ of the random noise that fixes the initial condition \cite{FootNote02}, the space
\begin{equation}
\label{eqn:CONTROLP}
\left\{ \alpha, \Delta\mu_o, J_m, I_r \right\},
\end{equation}
of our control parameters is in total 4-dimensional. We require two extra things: (i) an initial non uniform lattice that lifts the degeneracy between CB and CB$^*$ in
favor of the checkerboard, whose intensity is fixed at \mbox{$\delta\mu = -U_{NN}$}, and (ii) we isolate the rows of the lattice from tunneling to one another.
\begin{center}
\begin{figure}[h]
\includegraphics[width=0.9\linewidth]{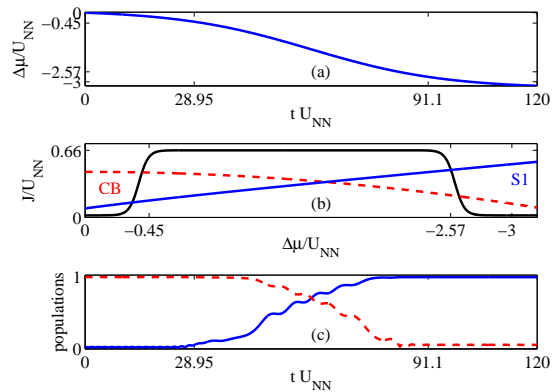}
\caption{(a) The pulse of local chemical potential as a function of time. (b) The step-like function is the tunneling coefficient as a function of $\Delta\mu$, while the thick (dashed) line is the 
tip of S1 (CB) insulating lobe. (c) Population inversion, from CB to S1 at the end of the process.
Notice the oscillation of populations when passing through the SF region of the phase diagram.}
\label{fig:DYNAMICS}
\end{figure}
\end{center}
In Fig. \ref{fig:DYNAMICS} we show the dynamics of the transferring process for $\alpha = 40\times 10^{-3} U_{NN}^2$, $\Delta\mu_o = -0.45U_{NN}$, $J_m = 0.66U_{NN}$ 
and $I_r = 4\times 10^{-3}$. 
In Fig. \ref{fig:DYNAMICS} (a) we plot the pulse of local chemical potential $\Delta\mu$ as a function of time. The smoothed step function in Fig. \ref{fig:DYNAMICS} (b) shows the tunneling coefficient as a function 
of $\Delta\mu$, while the dashed (thick) line is the tip of the CB (S1) insulating lobe. These pulses drive the population Fig. \ref{fig:DYNAMICS} (c) of S1 (thick line) to a steady value of
$P_{S1}(t_{fin}) = 0.992$ at the end of the process ($t_{fin} = 120/U_{NN}$), while the population of the CB state (dashed line) diminishes considerably. 
Notice the oscillation of populations when passing through the SF region of the phase diagram. Notice also that due to the definition 
(\ref{eqn:POPULATION}), populations do not have to sum up to one.

\begin{center}
\begin{figure}[h]
\includegraphics[width=0.9\linewidth]{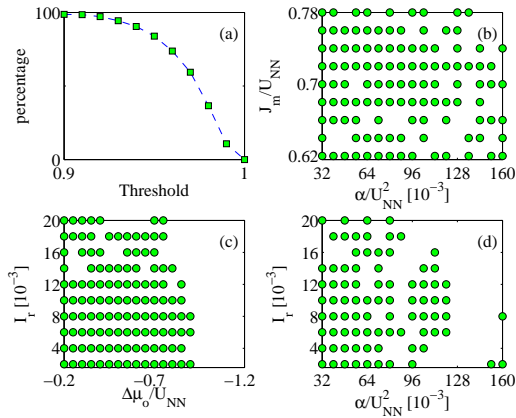}
\caption{Percentage of realizations (a) terminating with an S1 population bigger than threshold, versus threshold itself; as the threshold increases less realizations satisfy 
the required precision. (b-d) Slices of the discretize space of control parameters; the spots are for processes finishing in S1 with at least 0.98 population. In (b) we fix $I_r = 10\times 10^{-3}$ and  $\Delta\mu_o = 0.45U_{NN}$, in (c) and (d) we fix 
$\alpha = 40\times 10^{-3}U_{NN}^2$ and $\Delta\mu_o = -0.45U_{NN}$ respectively, for $J_m = 0.66U_{NN}$.}
\label{fig:STATISTICS}
\end{figure}
\end{center}

Having such a precise control on the parameters (\ref{eqn:CONTROLP}) is very challenging from the experimental point of view. Nevertheless, such a process is robust if
there is a reasonable range in which the parameters can vary without affecting the final result. The goal is of course the population of S1 to be as close as possible 
to 1 at the end of the process. We discretize the space of parameters (\ref{eqn:CONTROLP}) arbitrarily, and for every value of the parameters simulate the dynamics 
represented in Fig. \ref{fig:DYNAMICS}. The resulting statistics is shown in Fig. \ref{fig:STATISTICS}.
We have observed that there is a lower limit at $J_m = 0.6U_{NN}$, below which the transferring process does not work. For values of $J$ bigger than this limit, almost all 
the realizations end up in S1 state with a final population bigger than $0.8$, but the exact value depends on the control parameters of the single realization. 
In Fig.~\ref{fig:STATISTICS}~(a), we fix $J_m = 0.66U_{NN}$ and plot the percentage of the dynamics with $P_{S1}(t_{fin})$ coming through a given threshold as a function of the population threshold itself. As we increase the threshold the number of simulations ending up in S1 with a population that overcomes the given threshold decreases, up
to no simulations ending up at the ideal value $P_{S1}(t_{fin}) = 1$. This is a clear signature of a quantum controlled process. 
Notice however, that about 36$\%$ of our simulations terminate with S1 being populated at $0.98$. 

In Fig. \ref{fig:STATISTICS} (b,c,d) we show slices of the hypercube defined by the discretized space of control parameters (\ref{eqn:CONTROLP}), where the spots are placed in correspondence of 
the values giving a dynamics with $P_{S1}(t_{fin}) \geq 0.98$. In Fig.~\ref{fig:STATISTICS} (b) we fix $I_r = 10\times 10^{-3}$ and  $\Delta\mu_o = -0.45U_{NN}$, in Fig.~\ref{fig:STATISTICS} (c) and (d) we fix 
$\alpha = 40\times 10^{-3}U_{NN}^2$ and $\Delta\mu_o = -0.45U_{NN}$ respectively, for $J_m = 0.66U_{NN}$.
There is a closed region in the discretized $\left\{ \alpha,\Delta\mu_o,I_r \right\}$ space in which one always comes through the $0.98$ 
population threshold. This is true also for larger elementary cells, for which we have checked that the transferring process works the same way. This means that experimentally one has the freedom of setting the control parameters such that their small fluctuations 
do not affect the transfer process. This makes the specific process of population transferring from CB to S1 quite robust.  

\section{Trap effects}
\label{sec:FINITE}
So far we have considered an infinite lattice reproduced with a $4\times4$ or $2\times 2$ elementary cell with periodic boundary conditions. In real experiments atoms first are trapped in a 
harmonic trap and then the optical lattice is raised. Therefore it is important to understand the behavior of these systems in the  presence of a confining harmonic trap. 
Here we calculate the ground state of a finite $20\times20$ square lattice, where we superimpose a harmonic potential 
mimicked by local chemical potentials \cite{Jaksch,Troyer,Wessel}, without periodic boundary conditions. The range of the dipole-dipole interaction is cut at the fourth nearest neighbor as before. 
\begin{center}
\begin{figure}[h]
\includegraphics[width=1.0\linewidth]{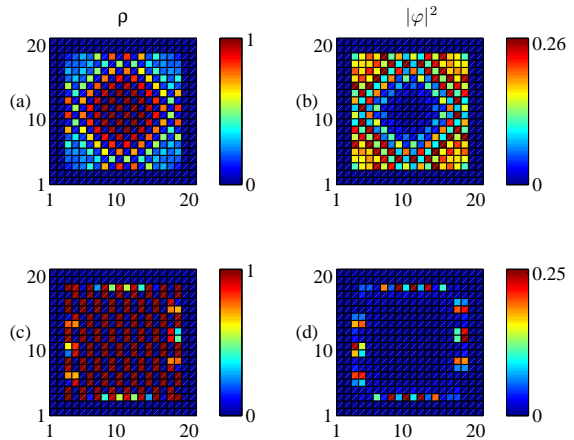}
\caption{(Color online) Density $\rho (x,y)$ and superfluid parameter $|\varphi (x,y)|^2$ in the harmonic trap.}
\label{CAKE}
\end{figure}
\end{center}
The harmonic potential for the system in Fig. \ref{CAKE} is $V(x,y) = \frac{K}{U_{NN}}\left[(x - x_0)^2 + (y - y_0)^2\right]$, where $(x_0,y_0)$ is the centre of 
the two dimensional isotropic trap.
The parameters for the system in Fig. (a,b) are $\mu/U_{NN} = 2.8$, $J/U_{NN} = 0.26$, and  $K = 107\times10^{-3} s^{-1}$ in units of $\hbar = 1$.
There is clearly a region around the center 
of the trap where the density $\rho(x,y)$ follows a checkerboard pattern and where the superfluid parameter $|\varphi(x,y)|^2$ is zero (see Fig. \ref{CAKE} (a,b)). Notice the supersolid-superfluid area that surrounds the Mott insulating phase.
In Fig. \ref{CAKE} (c,d) the parameters are $\mu/U_{NN} = 3.3$, $J/U_{NN} = 0.16$, and $K = 3.1\times10^{-3} s^{-1}$ in units of $\hbar = 1$. The density in the center of the trap (see Fig. \ref{CAKE} (c)) follows the metastable state atomic distribution of Fig. \ref{fig:TRAPLOBES} (II), with a zero 
superfluid parameter (see Fig. \ref{CAKE} (d)), while in the outer region of the trap a SF state is present.
In Fig. \ref{fig:TRAPLOBES} (a), the dashed and thick lines represent the extension of the previous trapping potentials, respectively for the (a,b) and (c,d) case.
\begin{center}
\begin{figure}[h]
\includegraphics[width=1.0\linewidth]{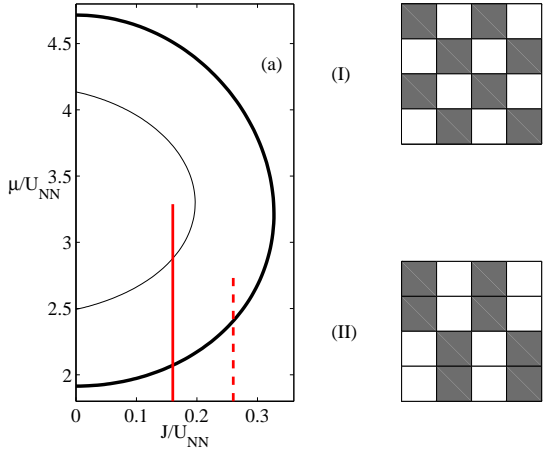}
\caption{(Color online) (a) The thick line is the lobe of the ground state (I), while the thin line represents the lobe of metastable state (II). Thick and dashed vertical
lines are the extension of the harmonic potential of Fig. \ref{CAKE} (c,d) and (a,b) respectively.}
\label{fig:TRAPLOBES}
\end{figure}
\end{center}

\section{Conclusion}
\label{sec:conclusion}
We have studied a single component gas of dipolar bosons in a two dimensional optical lattice. The atoms feature a polarized dipole moment perpendicular to the plane of the lattice
resulting in a long range interaction repulsive in every direction of the plane. 
The dipole-dipole interaction range has been truncated at the fourth nearest neighbor, and we have considered $4\times 4$ and $2\times 2$ unitary cells with periodic boundary conditions. 
We have shown that such a system possesses many almost degenerate metastable states often competing with the ground state.

We have studied the stability of these states and have shown that
the tunneling time scales exponentially with the number of sites
of the elementary cell of the corresponding
metastable configurations in the lattice, with a factor which depends
in a complicated way on the hopping parameter $J$, the energy difference
between the two metastable states, and the number of lattice sites involved in the
tunneling. In a previous work \cite{Menotti}, we also showed how to identify the state in the lattice through noise correlation measurements.

The mean field theory calculations have shown that, once the system is prepared in one of the metastable states, it is necessary to go into the superfluid region of the 
phase diagram in order to break the symmetry of the prepared state and transfer it to another one. Even though this is a quantum controlled process, very much
sensitive to the exact values of the control parameters during the dynamics, we have shown that the process is rather robust.

The capability of initializing, reading and manipulating these systems makes dipolar bosons in optical lattice very promising for applications in quantum information as quantum memories.

\section*{ACKNOWLEDGMENTS}
We would like to thank Mirta Rodriguez, Sibylle Braungardt, Luis Santos, Peter Zoller, Thierry Lahaye, and Tillman Pfau for useful discussions. We acknowledge financial support of ESF PESC ''QUDEDIS'', EU IP ''SCALA'', Spanish ''MICINN'' under 
Contract FIS 2005-04627 and Consolider-Ingenio 2010.
C.T. acknowledges the Fellowship Researcher in training (FI) “Supported by the Commission for Universities and Research of the Department of Innovation, Universities and Enterprises of the Catalan Government and the European Social Fund.” CM acknowledges
financial support by the EU through an EIF Marie-Curie Action.


\begin{thebibliography}{99}

\bibitem{review1} M. A. Baranov, \L. Dobrek, K Góral, L. Santos and M. Lewenstein, Phys. Scr. {\bf T102}, 74-81 (2002).

\bibitem{review2} C. Menotti, M. Lewenstein, arXiv:0711.3406.

\bibitem{review3} C. Menotti, M. Lewenstein, T. Lahaye, and T. Pfau, e-print arXiv:0711.3422.

\bibitem{review4} M.A. Baranov, to be published in Physics Reports 

\bibitem{pfau} A. Griesmaier, J. Werner, S. Hensler, J. Stuhler, and
T. Pfau, Phys. Rev. Lett. {\bf 94}, 160401 (2005); J. Stuhler,
A. Griesmaier, T. Koch, M. Fattori, T. Pfau, S. Giovanazzi, P. Pedri,
and L. Santos, Phys. Rev. Lett. {\bf 95}, 150406 (2005).

\bibitem{Lahaye} T. Lahaye, T. Koch, B. Fr\"ohlich, M. Fattori, J. Metz, A. Griesmaier, S. Giovanazzi, and T. Pfau, 
Nature {\bf 448}, 672 (2007).

\bibitem{EPJD} Special Issue {\it ``Ultracold Polar Molecules:
Formation and Collisions''}, Eur. Phys. J. D {\bf 31}, (2004).


\bibitem{review5} M. Lewenstein, A. Sanpera, V. Ahufinger, B. Damski, A. Sen(de), and U. Sen,
Adv. Phys. {\bf 56}, 243 (2007).

\bibitem{Fisher} M. P. A. Fisher, P. B. Weichman, G. Grinstein, and D. S. Fisher,
Phys. Rev. B. {\bf 40}, 546 (1989).

\bibitem{Jaksch} D. Jaksch, C. Bruder, J.I. Cirac, C.W. Gardiner, and P. Zoller,
Phys. Rev. Lett. {\bf 81}, 3108 (1998).


\bibitem{Bloch} M. Greiner, O. Mandel, T. Esslinger, T. W. H\"ansch, and I Bloch,
Nature  {\bf 415}, 39 (2002).

\bibitem{batrouni} G.G.~Batrouni and R.T.~Scalettar,
Phys. Rev. Lett. {\bf 84}, 1599 (2000).

\bibitem{sengupta} P. Sengupta, L.P.~Pryadko, F.~Alet, M.~Troyer, and
G.~Schmid, Phys. Rev. Lett. {\bf 94},207202 (2005).

\bibitem{sinha} D.L.~Kovrizhin, G. V. Pai, and S. Sinha, Europhys. Lett.
{\bf 72}, 162 (2005).

\bibitem{Goral} K. G\'oral, L. Santos, and M. Lewenstein, 
Phys. Rev. Lett.  {\bf 88}, 170406 (2002).

\bibitem{Menotti} C. Menotti, C. Trefzger, and M. Lewenstein,
Phys. Rev. Lett.  {\bf 98}, 235301 (2007).

\bibitem{Amit} Daniel J. Amit, Modeling Brain Function: The World Of Attractor Neural Networks 
(Cambridge University Press, Cambridge, 1989).

\bibitem{parisi} M. Mezard, G. Parisi, M. A. Virasoro, Spin Glass Theory And Beyond: An Introduction To The Replica Method And Its Applications (World Scientific, New Jersey, 2004).

\bibitem{polzik} K. S. Choi, H. Deng, J. Laurat, and H. J. Kimble, Nature  {\bf 452}, 67 (2008);
J. Appel, E. Figueroa, D. Korystov, M. Lobino, and A. I. Lvovsky, Phys. Rev. Lett. 100, 093602 (2008);
M. D. Eisaman, A. Andr\'e, F. Massou, M. Fleischhauer, A. S. Zibrov, and M. D. Lukin, Nature  {\bf 438}, 837 (2005);
T. Chaneli\'ere, D. N. Matsukevich, S. D. Jenkins, S.-Y. Lan, T. A. B. Kennedy, and A. Kuzmich, Nature  {\bf 438}, 833 (2005);
B. Julsgaard, J. Sherson, J. I. Cirac, J. Fiur\'a\v sek, and E. S. Polzik, Nature 432, 482 (2004).


\bibitem{sengupta2} J.-S.~Bernier, K.~Sengupta, and Y.B.~Kim, 
Phys. Rev. B {\bf 76}, 014502 (2007).

\bibitem{Jaksch01} D. Jaksch, V. Venturi, J. I. Cirac, C. J. Williams, and P. Zoller,
Phys. Rev. Lett. {\bf 89}, 040402 (2002).

\bibitem{Demler} D.-W. Wang, M. D. Lukin, and E. Demler,
Phys. Rev. Lett. {\bf 97}, 180413 (2006).

\bibitem{Sachdev} S. Sachdev, {Quantum Phase Transitions} (Cambridge University Press, Cambridge, 1999).

\bibitem{Wen} X.-G. Wen, {Quantum Field Theory Of Many Body Systems} (Oxford University Press, Oxford, 2004).

\bibitem{Perez} V. M. P\'erez-Garc\'ia, H. Michinel, J.I. Cirac, M. Lewenstein, and P. Zoller,
Phys. Rev. Lett. {\bf 77}, 5320 (1996).

\bibitem{Troyer} G. G. Batrouni, V. Rousseau, R. T. Scalettar, M. Rigol, A. Muramatsu, P. J. H. Denteneer, and M. Troyer,
Phys. Rev. Lett. Phys. Rev. Lett. {\bf 89}, 117203 (2002).

\bibitem{Wessel} S. Wessel, F. Alet, M. Troyer, and G. G. Batrouni,
Phys. Rev. A {\bf 70}, 053615 (2004).

\bibitem{FootNote01} In the MF we imply periodic boundary conditions.

\bibitem{FootNote02} Remember that we need a non zero order parameter to trigger 
non trivial dynamics, so we remove a random population from occupied sites of the CB and move it to empty ones.

\end{thebibliography}
\end{document}